\documentclass{desyproc}

\begin{document}
%------------------------------------
\title{Total cross section and elastic scattering from TEVATRON to LHC}

%for single authors the superscripts are optional
\author{{\slshape Hasko Stenzel}\\[1ex]
on behalf of the ATLAS Collaboration
Justus-Liebig Universit\"at Giessen, II. Physikalisches Institut\\
Heinrich-Buff Ring 16, D-35392 Giessen, Germany}

% if the proceedings are available online (e.g. at Indico)
% please enter the contribution ID or file_name below for the DOI
\contribID{32}
%\contribID{smith\_joe}

% TO THE CONFERENCE EDITORS: 
% please update the following information      
% before sending the template to the authors
\confID{1407}  % if the conference is on Indico uncomment this line
\desyproc{DESY-PROC-2009-03}
\acronym{PHOTON09} % if you want the Acronym in the page footer uncomment this line
\doi  % if there is an online version we will register DOIs

\maketitle

\begin{abstract}
A review of measurements of the total cross section in $p{\overline p}$ collisions at the TEVATRON 
and an outlook on the expected performance for similar determinations at the LHC is given.
The experimental method is based on the optical theorem to determine the total cross section independent 
of the machine luminosity. It consists of the extrapolation of the $t$-spectrum for elastic scattering to $t\to 0$ 
with a simultaneous measurement of the total inelastic rate. 
\end{abstract}

\section{Introduction}
One of the most elementary observables at hadron colliders is the total cross section. Precise measurements 
at high energies have been pioneered at the SPS by the UA4~\cite{UA4} collaboration. The method requires 
an instrumentation of the forward region far away from the interaction point but close enough to the beam  
to measure small-angle elastic scattering. At high energies the $t$-value for elastic $p{\bar p}$ and $pp$ scattering 
can be obtained from a measurement of the scattering angle $\theta$ by $-t=(p\cdot\theta)^2$ with the beam momentum $p$. 
The scattering angle is reconstructed from the proton impacts in the forward tracking devices, usually located in movable 
beam pockets, the so-called Roman Pots. On the other side, simultaneously with the elastic measurement, 
the total inelastic rate must be determined, which requires an extended pseudorapidity $\eta$ coverage. 
In this case the cross section is 
given by
\begin{displaymath}
\sigma_{\rm{tot}} = \frac{16 \pi}{1+\rho^2}\frac{\left. {\rm d}R_{\rm{el}}/{\rm d}t\right|_{t \to 0}}{R_{\rm{in}+}R_{\rm{el}}} \; , 
\, \rho=\left. \frac{Re(F_{\rm{el}}(t))}{Im(F_{\rm{el}}(t))} \right|_{t \to 0} \; ,
\end{displaymath}       
independent of the machine luminosity. In the following we shall review the measurements performed by the TEVATRON experiments and summarise the potential for the LHC experiments.

\section{Measurements at the TEVATRON}
\subsection{E710}
\begin{figure}[h!]
\centerline{\includegraphics[width=0.6\textwidth,angle=270]{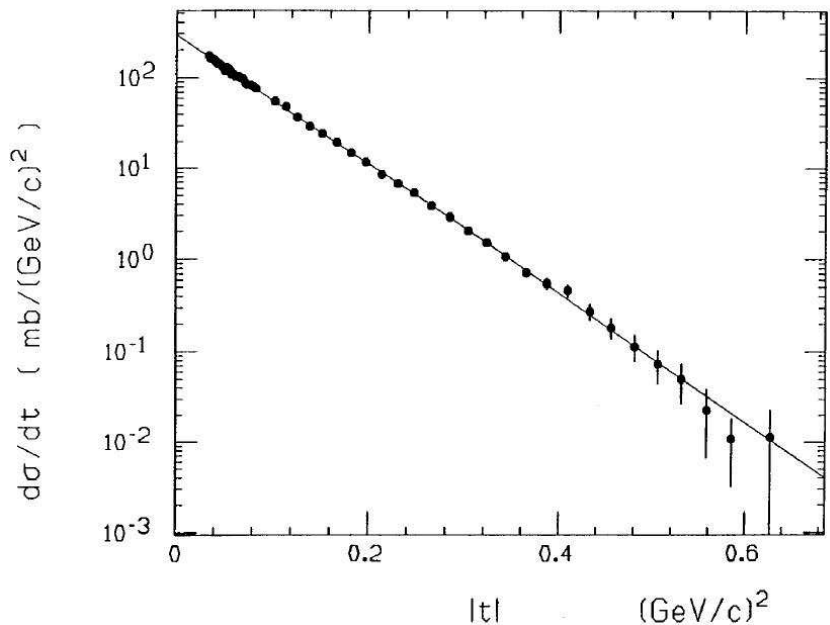}}
\caption{Elastic $t$ spectrum at 1.8 TeV measured by E710 \cite{E710_elastic}.}\label{fig:E710_elastic}
\end{figure}

At the FNAL TEVATRON $p{\overline p}$-accelerator the first total cross section measurement was done by the E710 experiment \cite{E710_elastic}-\cite{E710_rho} at $\sqrt{s}=1.8$ TeV. The E710 set-up was located around the E0 interaction point 
and was equipped with two Roman Pots at each side of E0 for the measurement of elastic protons. 
The Roman Pots were instrumented with wire chambers for the tracking and trigger scintillators. The locations 
of these detectors were in between 25 and 124 m from the interaction point. Under suitable beam conditions with a 
high $\beta^\star$ optics and low emittance the detector can be moved close to the beam at a distance of a few millimetres. In this case the impact position $u=x,y$ is related to the scattering angle by $u\propto \sqrt{\beta^\star\beta_{\rm{RP}}}\,\theta$, where $\beta$ is the betatron oscillation function resulting from the 
magnetic lattice. The measured $t$ spectrum from E710 is shown in fig.~\ref{fig:E710_elastic}.
The $t$-spectrum exhibits an approximately exponential fall-off with the nuclear slope $B$ in the 
considered range. The differential cross section is normally given by 
\begin{eqnarray}\label{eq:elastic}
\frac{{\rm d}\sigma}{{\rm d} t} & = & \frac{1}{16 \pi}\left |F_{\rm C}(t) + F_{\rm N}(t)\right |^2  \, \, ;\\
F_{\rm C}(t) & = & -8 \pi \alpha \hbar c \frac{G^2(t)}{t}\exp\left(i\alpha\Phi(t)\right) \, \, ;\nonumber \\
F_{\rm N} (t) & = & (\rho + i) \frac{\sigma_{\rm{tot}}}{\hbar c}\exp\left(\frac{-Bt}{2}\right) \, \, , \nonumber
\end{eqnarray}
with the Coulomb- and nuclear amplitudes $F_{\rm C}$ resp. $F_{\rm N}$, the electric form factor 
of the proton $G^2(t)$ and 
the Coulomb phase $\Phi(t)$. The Coulomb contribution is rather small in the $t$-range covered  
by E710, thus only the nuclear component was used to fit the spectrum (line in fig.~\ref{fig:E710_elastic}) 
in order to determine the optical point, which is defined by the value of the fit at $t=0$.  

\begin{figure}[h!]
\centerline{\includegraphics[width=0.8\textwidth,angle=180]{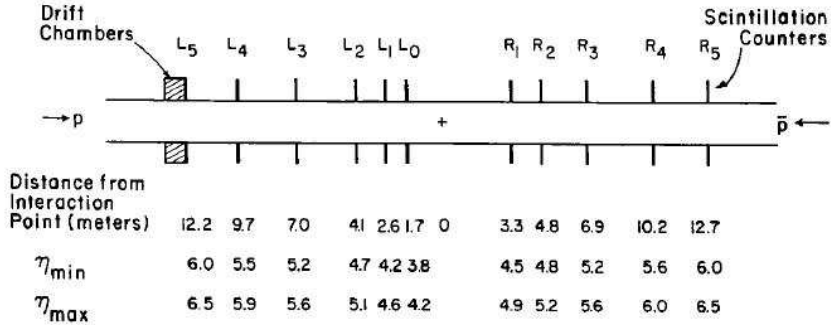}}
\caption{Inelastic detectors of E710.}\label{fig:E710_inelastic}
\end{figure}

The total inelastic rate was measured by the set-up depicted in fig.~\ref{fig:E710_inelastic}.  
It consisted of various annular scintillator counters segmented in four quadrants 
along the beam line covering a pseudo-rapidity range 
from 3.8 to 6.5, supplemented by drift chambers for precision tracking. With a trigger coincidence 
between the left and right arm about 70$\%$ of the inelastic interactions are covered. Some events, 
in particular from single diffraction processes escaped the left-right detection and were recovered 
by a single-arm OR trigger condition. This trigger is dominated by background from beam-gas interactions 
or halo protons but was measured using non-colliding bunches in the machine and then subtracted.  
A small fraction of events could still not be detected because of the limited angular acceptance coverage. 
The acceptance corrections were also obtained from the data using the drift chamber extrapolation the rate 
as function of the cosine of the maximum detected angle to zero, following a method developed by UA4~\cite{UA4}. 
The final results from E710 are summarised in Table~\ref{tab:tev}. 

\begin{table}[h!]
\caption{Results from TEVATRON experiments.}
\label{tab:tev}
\centerline{\begin{tabular}{|l|l|l|l|}
\hline
$\sigma_{\rm{tot}}$[mb]  &  $\sigma_{\rm{el}}$[mb] &  $\sigma_{\rm{tot}}/\sigma_{\rm{el}}$ & experiment\\\hline
80.03 $\pm$ 2.24    & 19.70 $\pm$ 0.85 & 0.246 $\pm$ 0.004 & CDF \\\hline
71.42 $\pm$ 2.41    & 15.79 $\pm$ 0.87 & 0.220 $\pm$ 0.008 & E811 \\\hline
72.8 $\pm$ 3.1    & 16.6 $\pm$ 1.6 & 0.230 $\pm$ 0.012 & E710 \\\hline
\end{tabular}}
\end{table}

\subsection{CDF}
\begin{figure}[h!]
\centerline{\includegraphics[width=0.6\textwidth,angle=0]{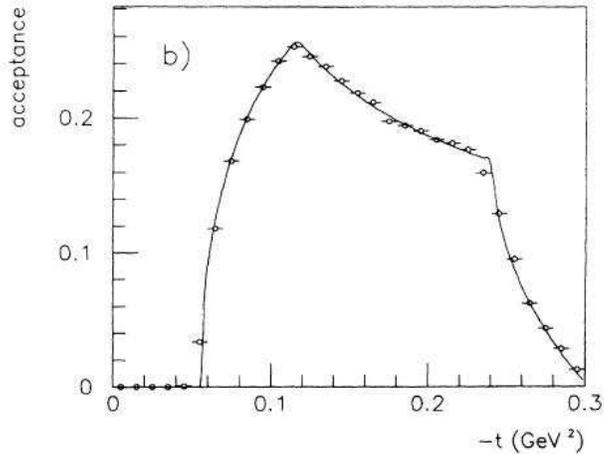}}
\caption{Acceptance of CDF for elastic scattering.}\label{fig:CDF_accept}
\end{figure}

The second experiment to perform measurements of elastic and total cross sections at the TEVATRON was CDF
\cite{CDF_elastic},\cite{CDF_total}. The elastic spectrometer of CDF consisted of three Roman Pot stations 
on the ${\bar p}$-side and two on the $p$-side, each instrumented with wire chambers, silicon strip detectors and 
trigger scintillators. The acceptance for elastic scattering was obtained from a detailed beam transport 
simulation, shown in fig.~\ref{fig:CDF_accept}.    
The inelastic measurement of CDF profited from the rich instrumentation around 
the interaction point, in particular the vertex TPC covering up to $|\eta|\leq 3.5$ allowed 
for a high-quality vertexing and enabled efficient background rejection. In the forward region 
a telescope from UA4 covering $3.8 \leq |\eta| \leq 5.5$ of drift chambers and a movable telescope 
enclosing completely the beam pipe when in beam position and covering then to $|\eta|\leq 7$ were installed. 
For the determination of the inelastic rate, in addition to the traditional left-right coincidence, a dedicated 
single-diffraction trigger consisting of a coincidence between the elastic spectrometer on one 
side detecting the diffractive proton and the opposite inelastic detectors tagging remnants of the diffractive dissociation was set up. Acceptance corrections for the limited angular coverage of the inelastic detectors and for 
losses of the diffractively scattered protons were obtained from detector and beam line simulations.
The result from CDF, given in Table~\ref{tab:tev}, is somewhat larger than the E710 result, both for the elastic and 
the total cross section. It should be noted that CDF has also provided a measurement~\cite{CDF_total} at 546 GeV, $\sigma_{\rm{tot}}=61.26 \pm 0.93$ mb, compatible with the UA4 result~\cite{UA4}. 
\subsection{E811}
\begin{figure}[h!]
\centerline{\includegraphics[width=0.5\textwidth, angle=270]{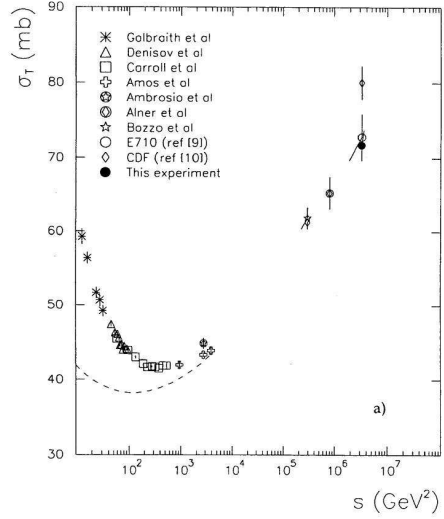}}
\caption{Summary of total cross section measurements at TEVATRON compared to lower energy data. 
{\em This experiment} is E811, taken from~\cite{E811_new}\label{fig:E811}}
\end{figure}

Given the discrepancy between 
E710 and CDF, another experiment E811~\cite{E811_det} was built upgrading the E710 location. E811 used the same 
inelastic detectors as E710 but replaced the elastic detectors by a scintillating fibre tracked coupled 
to a CCD read-out yielding a spatial resolution of 20 $\mu$m. The analysis technique was similar to the E710 
method but E811 accumulated larger statistics, extended the $t$-range and improved the systematic 
uncertainties~\cite{E811_new}. Their final result~\cite{E811_final} confirms the first E710 measurement and 
differs from CDF by 2.6 $\sigma$ on the total and by 2.9 $\sigma$ on the elastic cross section. 
The total cross section measurements are shown in fig.~\ref{fig:E811} together with lower energy $p\bar p$ results. 
   
\section{Prospect for measurements at LHC}
\subsection{TOTEM}
\begin{figure}[h!]
\centerline{\includegraphics[width=0.8\textwidth,angle=0]{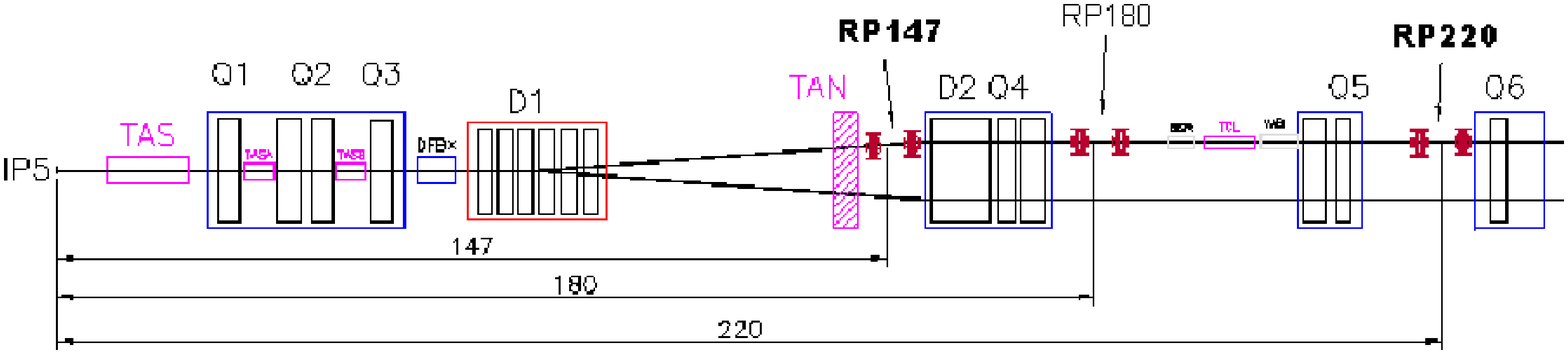}}
\caption{TOTEM elastic detectors.}\label{fig:TOTEM_elastic}
\end{figure}

At the LHC similar measurements are proposed by the TOTEM experiment~\cite{TOTEM_TDR},\cite{TOTEM_det} and ATLAS~\cite{ATLAS_det} with its dedicated ALFA subsystem~\cite{ATLAS_ALFA}.    
TOTEM shares the interaction point with CMS and has Roman Pot stations at 147 and 220 distance, with both detection 
in the vertical and horizontal planes to cover elastic scattering, diffraction and ease alignment. 
The elastic set-up is shown 
in fig.~\ref{fig:TOTEM_elastic}, each station houses silicon detectors for high-precision tracking and triggering.
At each location two stations separated by a few meters allow for the determination of the local track slope.
The detectors consist of 10 planes of silicon strips alternating in U- and V-Geometry, in a novel edgeless technology allowing to approach the beam as close as possible. The running scenario foresees a dedicated beam optics for 
the total cross section measurement with a $\beta^\star$ value of 1540m, in which case a good $t$-acceptance between 
0.002 and 1.5 GeV$^2$ is achieved, but also other configurations with smaller $\beta^\star$ are considered 
which opens up access to large $t$ values up to 10 GeV$^2$ and extended diffractive studies.     

\begin{figure}[h!]
\centerline{\includegraphics[width=0.8\textwidth,angle=0]{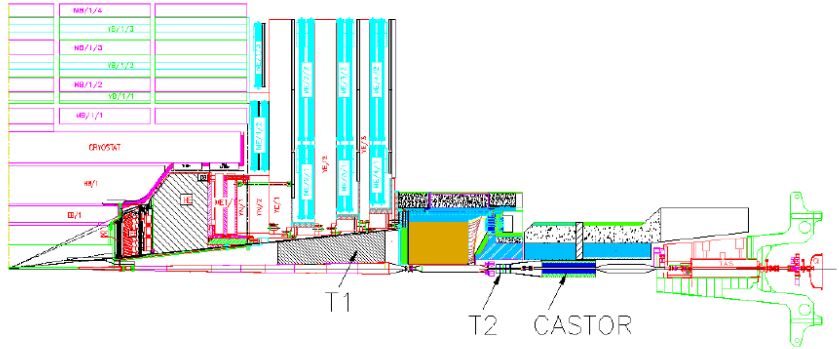}}
\caption{Inelastic detector T1 and T2 of TOTEM embedded in CMS components.}\label{fig:TOTEM_inelastic}
\end{figure}

The inelastic detectors are shown in fig.~\ref{fig:TOTEM_inelastic}. They consist of two tracking telescopes T1 and 
T2 in the pseudorapidity region 3.1 $\leq |\eta| \leq$ 6.5. T1 is of conical shape and inserted underneath the CMS 
flux return yoke at a distance between 7.5 and 10.5 m from the interaction point. Given the requirements on 
rate capabilities and radiation hardness, cathode strip chambers were selected as technology choice. The compact 
T2 telescope is installed in the forward shield of CMS in front of the CASTOR calorimeter at about 13.5 m 
from the interaction point. For T2 gaseous electron multiplier detectors were selected because of their excellent ageing 
performance. The telescopes will be used to trigger inelastic events with high efficiency and for the reconstruction 
of the primary vertex to discriminate against beam-gas and beam-halo background. In addition, integrated cross sections 
for hard and soft diffraction can be measured, also differential in $t$ and in the diffractive mass $M_{\rm x}$. 

The systematic uncertainties for the total cross section depend on the optics used, for the intermediate optics 
with $\beta^\star$=90 m an uncertainty of 4$\%$ is anticipated, dominated by the uncertainty of the optical functions 
needed for the $t$-reconstruction, but for the high $\beta^\star$ optics the uncertainty may decrease to 1$\%$. 
\subsection{ATLAS ALFA}
\begin{figure}[h!]
\centerline{\includegraphics[width=0.8\textwidth,angle=0]{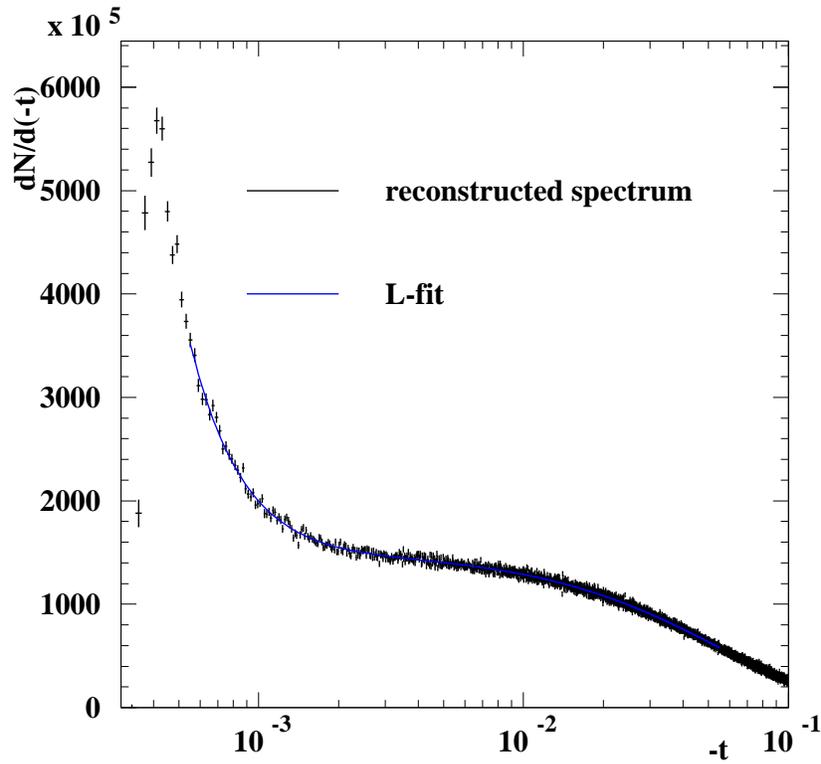}}
\caption{Simulated $t$-spectrum after reconstruction by the ATLAS ALFA detector compared to
the fit yielding the absolute luminosity calibration.}\label{fig:ALFA_elastic}
\end{figure}

The primary focus of the ATLAS ALFA detector is on the absolute luminosity calibration 
for the different relative luminosity monitors of ATLAS and specifically the dedicated relative monitor LUCID. Therefore the detector is designed to run only for a short 
period and under special beam conditions at low luminosity, thereby relaxing the constraints 
on the radiation tolerance. The luminosity can be determined according to eq.~\ref{eq:elastic} without using 
a measurement of the inelastic rate if the Coulomb-nuclear interference (CNI) region is covered. In this case the total cross 
section and the luminosity decouple because of the appearance of the Coulomb term. From a fit of the 
elastic $t$ spectrum the luminosity, total cross section, $B$ and $\rho$ parameters can be determined 
simultaneously. An example of a fit to the $t$-spectrum reconstructed with ALFA is shown in fig.~\ref{fig:ALFA_elastic}. 
This requires, however, that the CNI point where $|F_{\rm C}|\approx |F_{\rm N}|$, i.e. $t\approx 6.5\cdot10^{-4}$
 Gev$^2$ is reached.  It is therefore envisaged to use a very high $\beta^\star$ optics with 2600~m 
at a low luminosity of about $10^{27}$cm$^{-2}$s$^{-1}$ with low emittance. 
In this case the detector can approach the beam as close as 1.5mm (12 $\sigma$) and the 
CNI point is covered.

The ALFA detector is located at 240m from the ATLAS interaction point, at each location two 
vertical Roman Pot stations will be installed at a distance of 4 m from each other. The Roman 
Pots are instrumented with 10 planes of scintillating fibres, each double-sided plane carries 64 U- and 64 V-  
fibres. The fibres have a good sensitivity also at the edge oriented to the beam, which 
is essential for a high acceptance at small $t$. The scintillation light is recorded by multi-anode 
Photomultipliers. Since the detector is operated only for short periods it is foreseen to be removed 
after each period in order to prevent radiation damages during normal LHC operations. The scintillating fibre tracker 
is supplemented by plain scintillator tiles for triggering. The selected optics provides a parallel-to-point focusing 
in the vertical plane and accuracy of the $t$-scale depends crucially on an accurate vertical alignment. 
This is obtained by dedicated overlap detectors, which are made of horizontally aligned scintillating fibres housed 
in extrusions of the Roman Pots. When the Roman pots are in beam position the upper and lower overlap detector 
will measure the same halo tracks and the distance between the two pots can be determined with $\pm$ 10 $\mu$m precision.  

The elastic trigger will consist of a left-right coincidence requesting a signal from an upper detector on one 
side and lower on the other side in agreement with the back-to-back topology of elastic events, while 
coincidences of the same detector type can be used to measure the beam related background. Further background rejection 
can be achieved by using the local slope information from the two stations at each location. This 
enables a reconstruction of the transverse interaction vertex position through the beam transfer matrices. 

The systematic uncertainties are dominated by the knowledge of the optical functions, the detector alignment 
and the background determination. According to simulation a total uncertainty of 3$\%$ is achievable, 
for a running time of about 100 hours at a luminosity of $10^{27}$cm$^{-2}$s$^{-1}$.   

\section{Conclusion}
The long-standing tradition of total cross section measurements at hadron colliders from elastic scattering and 
the optical theorem has been successfully pursued at the TEVATRON with several measurements by 
CDF, E710 and E811 with an accuracy of about 3$\%$. At the LHC the dedicated TOTEM experiment 
will continue along these lines and even more precise measurements of 1$\%$ accuracy are expected, 
which will also be measured by ATLAS experiment with their dedicated ALFA system.  
 
\section{Acknowledgments}
This research was supported in part by the German Helmholtz Alliance ``Physics at the Terascale''.
\section{Bibliography} 
% ****************************************************************************
% BIBLIOGRAPHY AREA
% ****************************************************************************

\begin{footnotesize}
% IF YOU DO NOT USE BIBTEX, USE THE FOLLOWING SAMPLE SCHEME FOR THE REFERENCES
% ----------------------------------------------------------------------------

\end{footnotesize}

% ****************************************************************************
% END OF BIBLIOGRAPHY AREA
% ****************************************************************************

\end{document}